\documentstyle[11pt,newpasp,twoside,epsf]{article}
\begin{document}
%
%
\title{
The evolution of helium white dwarfs:\\
Applications for millisecond pulsars
}

\author{T.\ Driebe and T. Bl\"ocker}
\affil{Max-Planck-Institut f\"ur Radioastronomie, 
       Bonn, Germany}
\author{D. Sch\"onberner}
\affil{Astrophysikalisches Institut Potsdam, 
       Potsdam, Germany}
\section{White Dwarf Evolution}
Low-mass white dwarfs with helium cores (He-WDs) are known 
to result from mass loss and/or exchange events 
in binary systems where the donor is a low mass
star evolving along the red giant branch (RGB). 
Therefore, He-WDs are common components
in binary systems with either two white dwarfs
or with a white dwarf and a millisecond pulsar (MSP).
If the cooling behaviour of He-WDs is known from theoretical studies
(see Driebe et al.\ 1998, and references therein)
the ages of MSP systems can be calculated independently of the pulsar
properties provided the He-WD mass is known from spectroscopy.

\noindent
Driebe et al. (1998, 1999) investigated the evolution of He-WDs
in the mass range $0.18<M_{\rm WD}/{\rm M}_\odot<0.45$ 
using the code of Bl\"ocker (1995). The evolution of a $1\,{\rm M}_\odot$-model was
calculated up to the tip of the RGB.
High mass loss terminated the RGB evolution at appropriate positions depending
on the desired final white dwarf mass.
When the model started to leave the RGB, mass loss was virtually switched off
and the models evolved towards the white dwarf cooling branch.
The applied procedure mimicks the mass transfer in binary systems.

\noindent
Contrary to the more massive C/O-WDs ($M_{\rm WD}\ga 0.5\,{\rm M}_\odot$, carbon/oxygen core), 
whose progenitors have also evolved through the asymptotic giant branch phase, 
He-WDs can continue to burn hydrogen via the pp cycle along
the cooling branch down to very low effective
temperatures, resulting in cooling ages of the order of Gyr, \hbox{i.\ e.} of the 
same order of magnitude as the spin-down ages of millisecond pulsars.
The mass-radius-relation for He-WDs shows significant evolutionary effects 
due to this residual nuclear burning.
The ongoing hydrogen burning in He-WDs is provided by large envelope masses
which decrease with increasing  $M_{\rm WD}$ (see Bl\"ocker et al. 1997). 
In contrast to C/O-WDs, He-WDs cool down the slower the smaller their mass.
\section{Application to millisecond pulsars}
\noindent
Based on their evolutionary models 
Driebe et al.\ (1998) determined the mass of the He-WD
companion of the MSP PSR J1012+5307 (Nicastro et al., 1995, see also Sarna et al.\ 1998).
They found $M_{\rm WD}=0.19\pm 0.02\,{\rm M}_{\odot}$ and $0.15\pm 0.02\,{\rm M}_{\odot}$
using the spectroscopic data of van Kerkwijk et al.\ (1996) and Callanan et al.\ (1998), resp.
Thus, the mass ratio $M_{\rm pulsar}/M_{\rm He-WD} \approx 9.5\pm 0.3$ 
(van Kerkwijk et al.\ 1998, priv.\ comm.) gives a 
pulsar mass $M_{\rm pulsar}=1.43\pm 0.25\,{\rm M}_{\odot}$ and $1.81\pm 0.25\,{\rm M}_{\odot}$, resp.
From our models we derived a white dwarf age of \hbox{$\approx 6$ Gyr} in excellent 
agreement with the pulsar's spin-down age of 7 Gyr.

\begin{figure}
\plotfiddle{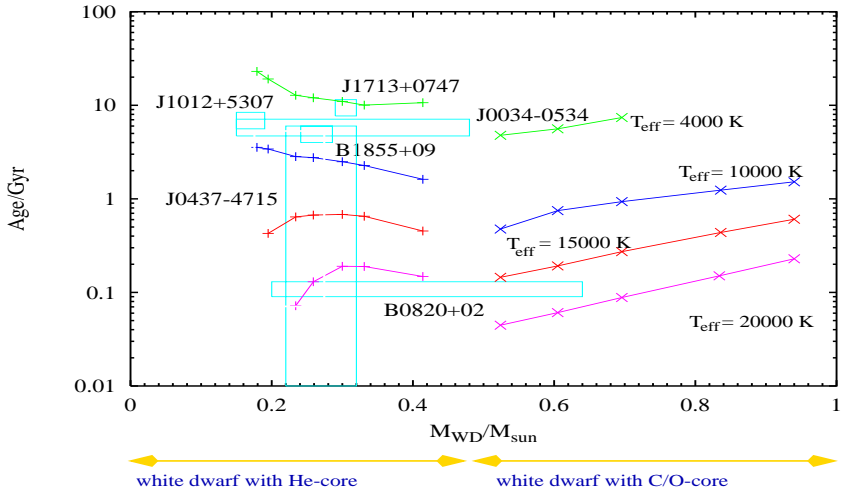}{8cm}{270}{45}{35}{-200pt}{260pt}
\vspace{-20mm}
\caption{\label{figmsp}
Ages of 6 MSP systems based on pulsar measurements as a function of
the WD companion mass (see \hbox{Burderi et al.\ 1998).} The boxes
refer to the uncertainties of spin-down ages and mass determinations.
Cooling ages for evolutionary He- and C/O-WD models 
(Driebe et al.\ 1998, Bl\"ocker 1995) are given for four effective temperatures.
}
\end{figure}

\noindent
We studied other MSP systems as well assuming the correspondence of white dwarf 
cooling age and $\tau_{\rm spin-down}$. Selecting only systems with well given 
ages and/or masses, the white dwarf effective temperatures and surface gravities 
can be determined with the present evolutionary models. \hbox{Figure \ref{figmsp}}
shows the results for 6 MSP systems.
Note that the consideration of fully evolutionary He-WD models is crucial 
for the determination of the effective temperature.
More details will be given in Sch\"onberner et al. (1999). 

\begin{references}
\reference  Bl\"ocker, T. 1995, \aap\, 297, 727
\reference  Bl\"ocker, T. et al.\ 1997,
            in: {\it White Dwarfs},\\  J. Isern, M. Hernanz, E. Garcia-Berro (eds.),
            Kluwer, Dordrecht, p.\ 57
\reference  Burderi, L., King, A. R., Wynn, G. A. 1998, \mnras\, 300, 1127
\reference  Callanan, P. J., Garnavich, P. M., Koester, D. 1998, \mnras\, 298, 207
\reference  Driebe, T., Sch\"onberner, D., Bl\"ocker, T., Herwig, F. 1998, \aap\, 339, 123
\reference  Driebe, T., Bl\"ocker, T., Sch\"onberner, D., Herwig, F. 1999, \aap\, 350, 89
\reference  van Kerkwijk, M. H., Bergeron, P., Kulkarni, S. R. 1996, \apj\, 467, L89
\reference  Nicastro, L. et al.\ 1995, \mnras\, 273, L68
\reference  Sarna, M. J., Antipova, J., Muslimov, A. 1998, \apj\, 499, 407
\reference  Sch\"onberner, D., Driebe, T., Bl\"ocker, T., 1999, A\&A, submitted
\end{references}
\end{document}